\documentclass[12pt]{iopart}
\usepackage{graphicx}

\newcommand{\be}{\begin{equation}}
\newcommand{\ee}{\end{equation}}
\newcommand{\bea}{\begin{eqnarray}}
\newcommand{\eea}{\end{eqnarray}}

\def\pmb#1{\setbox0=\hbox{$#1$}%
  \kern-.025em\copy0\kern-\wd0
  \kern.05em\copy0\kern-\wd0
  \kern-.025em\raise.0433em\box0}

\def\alt{\mathrel{\hbox{\rlap{\hbox{\lower4pt\hbox{$\sim$}}}\hbox{$<$}}}}
\begin{document}
\title{An experiment to measure electromagnetic memory}

\author{Lydia Bieri}
\address{Dept. of Mathematics, University of Michigan, Ann Arbor, MI 48109-1120, USA}
\ead{lbieri@umich.edu}
\author{David Garfinkle}
\address{Dept. of Physics, Oakland University,
Rochester, MI 48309, USA}
\address{and Michigan Center for Theoretical Physics, Randall Laboratory of Physics, University of Michigan, Ann Arbor, MI 48109-1120, USA}
\ead{garfinkl@oakland.edu}


\date{\today}

\begin{abstract}
We describe an experiment to measure the electromagnetic analog of gravitational wave memory, the so-called electromagnetic memory.  Whereas gravitational wave memory is a residual displacement of test masses, electromagnetic memory is a residual velocity (i.e. kick) of test charges. 

The source of gravitational wave memory is energy that is not confined to any bounded spatial region: in the case of binary black hole mergers the emitted energy of gravitational radiation as well as the recoil energy of the final black hole.  Similarly, electromagnetic memory requires a source whose charges are not confined to any bounded spatial region.  

While particle beams can provide unbounded charges, their currents are too small to be practical for such an experiment.  Instead we propose a short microwave pulse applied to the center of a long dipole antenna.  In this way the measurement of the kick can be done quickly enough that the finite size of the antenna does not come into play and it acts for our purposes the same as if it were an infinite antenna.

\end{abstract}


\maketitle

\section{Introduction}
\label{intro}
\indent

\subsection{Gravitational Wave Memory}

The gravitational wave memory effect was first computed by Ya.B. Zel'dovich and A.G. Polnarev in 1974 in the linearized theory \cite{zeldovichpolnarev}. They studied radiation of gravitational waves from a cluster of superdense stars, and they computed the memory resulting from a flyby of such stars. In 1991, D. Christodoulou \cite{chrmemory} investigated the fully nonlinear theory and computed the memory resulting from the emitted energy in gravitational radiation, deriving a new contribution to the memory. 
Thus, the latter is caused by the waves themselves radiating away energy. 
In hindsight it has become clear that the main difference between the two types of memory is that the Zel'dovich-Polnarev contribution is sourced by the difference of initial and final masses and velocities of the source (which we call ordinary memory), while the Christodoulou contribution (which we call null memory) is due to the total energy radiated away along null hypersurfaces. In fact, the sources for ordinary memory travel at speeds slower than light, whereas the sources for the null memory travel at the speed of light. 
Zel'dovich and Polnarev \cite{zeldovichpolnarev} considered gravitational waves for weak fields and small velocities, which allowed them to use the quadrupole approximation computing the memory as an overall change of the second time derivative of the source's quadrupole moment. On the other hand, Christodoulou \cite{chrmemory} investigated gravitational waves for strong fields and high velocities in the fully nonlinear theory. The null contribution emerges from the energy density (Bondi news) $|\Xi|^2$ integrated over retarded time. Thus, it is a cumulative effect building up during the passage of the wave packet. 

The pioneering works \cite{zeldovichpolnarev} and \cite{chrmemory} were followed by several contributions \cite{blda1, blda2, braginsky, braginskyg, will, thorne, thorne2, jorg}. Some physical forms of matter contribute to the null memory when coupled to the Einstein field equations \cite{lbdg3}. This is true in particular for the Einstein-Maxwell system \cite{1lpst1}, \cite{1lpst2}, as well as for neutrino radiation (where the neutrinos are approximated as massless) \cite{neutrinos} as it occurs in a core-collapse supernova or a binary neutron star merger. 

To understand the nature of gravitational wave memory, it is helpful to think in terms of the Weyl tensor and the geodesic deviation equation \cite{lbdg3}. The residual relative displacement of test masses is simply two integrals with respect to time of the Weyl tensor.  Memory is then a consequence of the Bianchi identities viewed as equations of motion for the Weyl tensor, with the sources of memory simply the right hand sides of those equations of motion.

\subsection{Electromagnetic Memory}

Thinking of memory in this way, it became clear to us \cite{lbdg2} that there must be an electromagnetic analog of gravitational wave memory: (i) the electric and magnetic fields acting as analogs of the electric and magnetic parts of the Weyl tensor, (ii) Maxwell's equations acting as analogs of the Bianchi identities, and (iii) one integral of the electric field acting as the analog of two integrals of the Weyl tensor.  Since a charge $q$ with mass $m$ obeys the equation of motion 
\begin{equation}  
m {\frac {{d^2} {\vec x}} {d{t^2}}} = q {\vec E} \; ,
\end{equation}
once all electromagnetic waves have passed the charge has received a kick given by
\begin{equation}
\Delta {\vec v} = {\frac q m} {\int _{-\infty} ^\infty} {\vec E} d t
\label{kick}
\end{equation}
Thus this kick is the electromagnetic analog of the residual relative displacement of gravitational wave memory.

Just as in the gravitational case, there are two types of electromagnetic memory \cite{lbdg2}: an ordinary memory sourced by charges traveling slower than light and a null memory sourced by charges traveling at the speed of light. 
In other words, the ordinary memory is due to a change of the radial component $E_r$ of the electric field over time, and the null EM memory is due to the charge radiated to infinity. 
Unlike the gravitational case, nature does not have electric charges that travel at the speed of light. 
However, it was shown in \cite{lbatdgbw} that massive particles exhibit the ordinary memory effect, but will also mimic the null memory when they have high velocities. 
Furthermore, null memory is a mathematical property of the Maxwell-Klein Gordon equation in the case where the Klein-Gordon field is massless \cite{MKG}. 

There has also been a growing interest in memory analogues in quantum theories. See some of the more recent articles cited above for more references.

\subsection{Measuring Gravitational Wave Memory and Electromagnetic Memory}

So far (at the time of the writing of this article) no memory has yet been measured (nor recognised as such in a measurement) in any physical theory. The detection of memory will mark an important breakthrough. 

For most ``regular" sources of gravitational radiation in general relativity (GR), a kick like in the electromagnetic (EM) memory is forbidden by the Einstein equations. However, this is not the case, and new memory arises for spacetimes of slow fall-off towards infinity, as was shown in \cite{lydia4}, \cite{lydia14}. 
A special situation was studied by L.P. Grishchuk and A.G Polnarev \cite{GrPo1}, where test masses get a kick. This is the velocity-coded memory. Moreover, a forthcoming paper \cite{BieriPolnarev1} derives velocity-coded memory for a specific astrophysical scenario. 

The GR displacement memory manifests itself as a permanent change of the spacetime, showing as a permanent displacement of test masses in a detector like LIGO and as a frequency change of pulsars' pulses in NANOGrav.

Measuring gravitational wave memory is a goal of current gravitational wave experiments \cite{Lasky1}, \cite{gwmemmeasure} and will mark an important breakthrough in science. Once this goal is achieved, memory will be an important tool to obtain more insights into the physics of gravitational wave sources. However, this goal is difficult to attain.  Since electromagnetism is a much stronger force than gravity, electromagnetic memory should be much easier to measure than gravitational wave memory.  Indeed since so much of modern technology is based on electromagnetism and electromagnetic waves it is at first very surprising that electromagnetic memory has not yet been measured.  

The reason that it hasn't been measured yet is that our devices for making electromagnetic waves have their charges and currents confined to a finite region of space, whereas electromagnetic memory (like gravitational wave memory) requires sources that are not confined to any finite region.  

Despite this difficulty, it should be possible to devise a strategy to measure electromagnetic memory.  The rest of this paper will be devoted to setting out such a strategy.

\section{Measuring the EM Memory}
\label{measure}

\subsection{Strategy}
\label{strategy}

Since EM memory is sourced by unbound charges, one possible strategy to measure EM memory is to use actual unbound charges, such as those in the beams of particle accelerators.  However, particle beams tend to have low currents, so this would not be a practical strategy.

Instead, we suggest {\emph {to create a situation of unbound charges for a short time in an antenna}}. These circumstances have to be maintained for at least as long that a measurement can be done. 
Our idea is to use a long dipole antenna to create a pulse of charges that act {\emph {for the purpose of the experiment}} as though they are unbound.  Here ``for the purpose of the experiment'' means that the measurement can be accomplished in a sufficiently short time that the event where the charges reach the end of the antenna is not within the past light cone of the event where the measuring apparatus ceases taking data.  For this, we will need a very short electromagnetic pulse applied to the center of the antenna so that the effects of the end of the pulse reach the measuring apparatus before the effects of the charges from the beginning of the pulse reaching the end of the antenna.

What sort of apparatus should be used to measure the memory?  In principle one could simply use a test charge that starts at rest and measure the kick that it receives.  However, in practice it will probably be easier to simply measure the electric field at a particular position in the radiation zone and note that its integral is nonzero. To do so, one could use another dipole antenna set up appropriately in the far field region. Before the charges reach the ends of the dipole antenna of the source, they induce a current in the receiver antenna due to the fact that the integral of the electric field over time is nonzero. 
Once the charges get reflected at the ends of the source antenna, this will destroy the memory. 
Consequently, the measurement has to be done within a short time before the pulse reaches the end of the antenna.

\subsection{Details of the Experiment}
\label{details}

Having outlined the general strategy, we now turn to a more detailed description of the proposed experiment. 
In the following, the current density ${\vec J}$ will be in the $z$ direction, that is we have the current density vector $\vec{J} = J \hat{z}$. 
The dipole antenna is driven at its center by a time dependent current $I(t)$.  This induces a pulse of current that travels at speed $v$ in each arm of the antenna, so that the current density is given by
\be
{\vec J}(t,{\vec x}) = I \left ( t - {\frac {|z|}  v}\right ) \delta (x) \delta (y) {\hat z}
\label{Jeqn}
\ee
From ${\partial _t} \rho + {\vec \nabla}\cdot {\vec J}=0$ it then follows that the charge density is 
\be
\rho(t,{\vec x}) = {\frac {\pm 1}  v} I \left ( t - {\frac {|z|}  v}\right ) \delta (x) \delta (y) 
\label{rhoeqn}
\ee
(where the plus sign goes with positive $z$). The retarded expressions for the scalar and vector potentials are \cite{Jackson}
\bea
\Phi (t,{\vec x}) = {\frac 1  {4 \pi {\epsilon _0}}} \int {d^3} {x'} {\frac {\rho ({t'},{{\vec x}'})}  {|{\vec x} - {{\vec x}'}|}}
\label{Phiret}
\\
{\vec A} (t,{\vec x}) = {\frac 1  {4 \pi {\epsilon _0}{c^2}}} \int {d^3} {x'} {\frac {{\vec J} ({t'},{{\vec x}'})}  {|{\vec x} - {{\vec x}'}|}}
\label{Aret}
\eea
where the retarded time is given by ${t'} = t - |{\vec x} - {{\vec x}'}|/c$
Using equations (\ref{rhoeqn}) and (\ref{Jeqn}) in equations (\ref{Phiret}) and (\ref{Aret}) we obtain
\bea
\Phi (t,{\vec x}) = {\frac 1  {4 \pi {\epsilon _0}}} \; {\frac 1  v} \left [
- {\int _{-\infty} ^ 0} d {z'} {w ^{-1}} {I_+} + {\int _0 ^\infty} d {z'} {w ^{-1}} {I_-} \right ]
\label{Phidipole}
\\
{\vec A} (t,{\vec x}) = {\frac 1  {4 \pi {\epsilon _0}}} \; {{\hat z}\over {c^2}} \left [
 {\int _{-\infty} ^ 0} d {z'} {w ^{-1}} {I_+} + {\int _0 ^\infty} d {z'} {w ^{-1}} {I_-} \right ]
\label{Adipole}
\eea
where the quantities $w$ and ${I_\pm}$ are given by
\bea
w \equiv {\sqrt {{r^2} {\sin ^2}\theta +{{(z-{z'})}^2}}}
\\
{I_\pm} \equiv I \left ( t - {\frac w  c} \pm {\frac {z'} v} \right )
\eea
Since the configuration is axisymmetric, the electric field in spherical coordinates takes the form ${\vec E} = {E_r} {\hat r} + {E_\theta} {\hat \theta}$.  Applying the expression ${\vec E} = - {\partial _t}{\vec A} - {\vec \nabla } \Phi$ to equations (\ref{Phidipole}-\ref{Adipole}) some straightforward but tedious algebra (involving integration by parts) yields the following:
\bea
{E_\theta} &=& {\frac 1  {4 \pi {\epsilon _0}}} \sin \theta \biggl [ {\frac {I(u)}  r} {\frac {2 v}  {{c^2} - {v^2}{\cos^2}\theta}}
\nonumber
\\
&+& \left ( {\frac 1  {v^2}} - {\frac 1  {c^2}} \right )  {\int _{-\infty} ^0} d{z'} {w^{-1}}{I_+} {{\left ( {\frac {z-{z'}} c} + {\frac w  v} \right ) }^{-2}} \left (  {\frac w  c} - {\frac {z'}  v} \right ) 
\nonumber
\\
&+& \left ( {\frac 1  {v^2}} - {\frac 1  {c^2}} \right )  {\int _0 ^\infty}  d{z'} {w^{-1}}{I_-} {{\left ( {\frac {z-{z'}} c} - {\frac w  v} \right ) }^{-2}} \left (  {\frac w  c} + {\frac {z'}  v} \right ) \biggr ]
\label{Etheta}
\eea
\bea
{E_r} = {\frac 1  {4 \pi {\epsilon _0}}}  \left ( {\frac 1  {v^2}} - {\frac 1  {c^2}} \right ) \biggl [ {\int _{-\infty} ^0} d{z'} {w^{-1}}{I_+} {{\left ( {\frac {z-{z'}} c} + {\frac w  v} \right ) }^{-2}} \left ( - {\frac {w\cos \theta}  c} + {\frac { {z'}\cos \theta -r}  v} \right ) 
\nonumber
\\
+ {\int _0 ^\infty}  d{z'} {w^{-1}}{I_-} {{\left ( {\frac {z-{z'}} c} - {\frac w  v} \right ) }^{-2}} \left ( - {\frac {w \cos \theta}  c} + {\frac {r-{z'}\cos \theta }  v} \right ) \biggr ]
\label{Er}
\eea
Here $u=t-r/c$.

These expressions vastly simplify in the radiation zone, since we keep only terms to order $r^{-1}$.  To this order we have ${E_r}=0$ and 
\be
{E_\theta} = {\frac 1  {4 \pi {\epsilon _0}}} \sin \theta {\frac {I(u)}  r} {\frac {2 v}  {{c^2} - {v^2}{\cos^2}\theta}}
\ee

It then follows immediately that the electromagnetic memory is
\be
{\int _{-\infty} ^\infty} {\vec E} d t = {\hat \theta} {\frac {\sin \theta }  r}  {\frac 1  {4 \pi {\epsilon _0}}} {\frac {2 v}  {{c^2} - {v^2}{\cos^2}\theta}}
{\int _{-\infty} ^\infty} \, I(u) \, du 
\ee
The physical interpretation of this result is straightforward: ${\int _{-\infty} ^\infty} I(u) du$ is the amount of charge moving along the positive $z$ axis, with an opposite charge moving along the negative $z$ axis.  Thus the dipole antenna has a linearly growing dipole moment which is the source of the electromagnetic memory.

Though memory is a phenomenon of the radiation zone, a practical experiment will not have the luxury of going to arbitrarily far distances.  We now consider an example pulse and use equations (\ref{Etheta}-\ref{Er}) to find the behavior of the electric field at reasonable laboratory distances.  For simplicity we confine ourselves to the equatorial plane and use units where $c=1$.  Then ${E_r}=0$ and 
\bea
{E_\theta} = {\frac 1  {4 \pi {\epsilon _0}}} \biggl [ {\frac {2 v I(u)}  r} 
\nonumber
\\
+ \left ( {\frac 1  {v^2}} - 1 \right )  {\int _{-\infty} ^\infty} d{z'} {w^{-1}} {{\left ( {\frac w  v} + |{z'}| \right ) }^{-2}} \left ( w + {\frac {|{z'}|}   v} \right ) I \left ( t - w - {\frac {|{z'}|} v} \right )
 \biggr ]
\label{Etheta2}
\eea

Maxwell's equations are scale invariant, but a practical experiment should take place at a reasonable laboratory scale.  In our example we will take the unit of time to be the nanosecond.  Since we have units where $c=1$ and since $c$ in mks units is $c=3.00 \times {{10}^8}$ m/s, it follows that our unit of length is 0.3 m.  

\begin{figure}
    \includegraphics[width=0.7\linewidth]{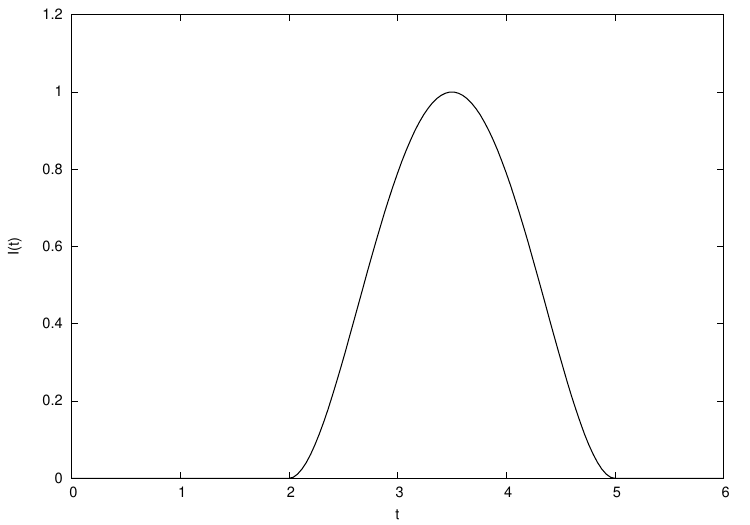}
    \caption{current applied to the center of the dipole antenna}
    \label{currentfig}{}
\end{figure}

Because Maxwell's equations are linear, we can pick an arbitrary scale for the pulse of current, with our results to be rescaled to fit the parameters of the experiment.  We pick the scale so that $I_{\rm {max}}/(4\pi {\epsilon_0})=1$ where $I_{\rm {max}}$ is the maximum value of the current.  We pick a time interval $({t_1},{t_2})$ and choose the pulse to vanish outside the interval and to take the form
\be
I(t)= 4 \pi {\epsilon _0} {{\left [{\frac {4(t-{t_1})({t_2}-t)} {{({t_2}-{t_1})}^2}} \right ]}^2}
\ee
within the interval. Figure \ref{currentfig} shows a graph of $I(t)$ (in units of $4\pi {\epsilon _0}$) vs. $t$ for 
${t_1}=2$ and ${t_2}=5$.

\begin{figure}
    \includegraphics[width=0.7\linewidth]{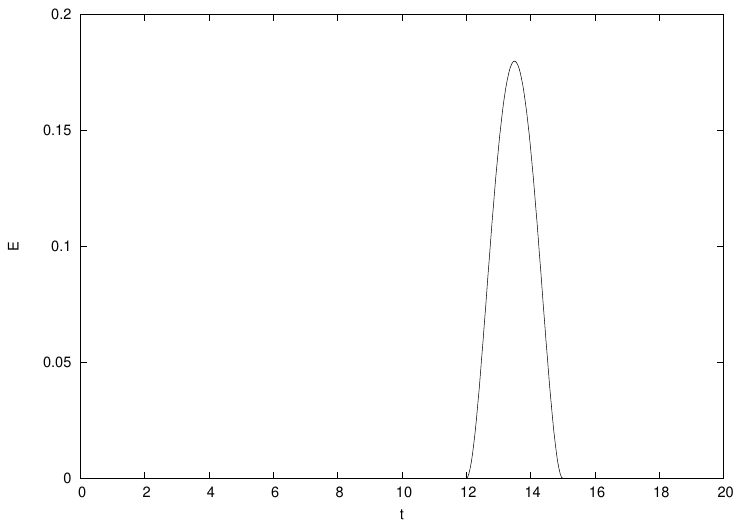}
    \caption{electric field at $r=10$ in the equatorial plane}
    \label{elecfig}{}
\end{figure}

Next we pick an equatorial observer at a distance of 10 in our units ({\it i.e.} 3 m).  We choose $v=0.9$ and use equation (\ref{Etheta2}) to find the electric field at the position of this observer as a function of time.  Here we evaluate the integrals in equations (\ref{Etheta2}) numerically. The results are plotted in figure (\ref{elecfig}).  

\begin{figure}
    \includegraphics[width=0.7\linewidth]{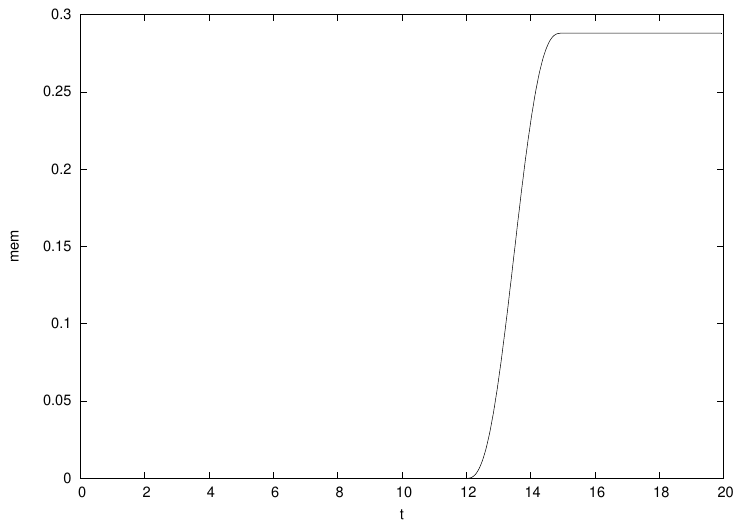}
    \caption{memory (integral of the electric field) at $r=10$ in the equatorial plane}
    \label{memfig}{}
\end{figure}

Note that one can tell from the form of the graph that the integral of the tangential component of the electric field (the memory) will be nonzero.  To make this more precise, we plot in figure \ref{memfig} the integral of the tangential component of the electric field.

Note that the measurement of memory can be accomplished by time 15, so if we stop the measurement at that time, each arm of the antenna needs to be long enough so that the event where the beginning of the pulse reaches the end of the antenna is not in the past light cone of the event at which the measurement ends.  Let $L$ be the length of each arm of the antenna and Let $t_m$ be the time at which the measurement ends and $r$ be the distance from the origin of the point in the equatorial plane at which the measurement is made.  Then in Cartesian coordinates $(t,x,y,z)$ the event where the measurement ends is $=({t_m},r\cos \phi,r\sin \phi,0)$ while the event where the beginning of the pulse reaches the end of the antenna is $({t_1}+L/v,0,0,L)$.  The condition that the second event is not in the past light cone of the first becomes the condition $L>{L_{\rm {min}}}$ where
\be
{L_{\rm {min}}} = {\frac v  {1-{v^2}}} \left [ {t_m}-{t_1} - {\sqrt { {v^2}{{({t_m}-{t_1})}^2} +(1-{v^2}){r^2}}} \right ]
\ee
Since in our case $v=0.9,\, {t_1}=2, \, {t_m}=15$ and $r=10$ we find that the condition becomes $L>2.44$ in our units or $L> 0.732 {\rm m} $ in mks units.

\section{Conclusion} 
\label{conclusion}

Memory measurement is upon us. Whether the first measurement will be done in GR or EM, will be seen. More importantly, besides confirming theoretical predictions, memory will serve as an important tool to explore the physics of the sources. Moreover, measuring EM memory may serve as a first step to investigate further memory analogues in quantum theories. 

It should be straightforward to measure electromagnetic memory using the methods sketched out in this paper.  Certainly electromagnetic memory will be far easier to detect than gravitational wave memory.  We encourage all experimentalists with an interest in memory and with access to the appropriate equipment to make this measurement.

\section*{Acknowledgements}

DG was supported by NSF grant PHY-2102914 to Oakland University.  LB was supported by NSF grant DMS-2204182 to the University of Michigan.

\end{document}